# Optical frequency comb Fourier transform spectroscopy of the CH$_2$$^{79}$Br$^{81}$Br, CH$_2$$^{79}$Br$_2$, and CH$_2$$^{81}$Br$_2$ isotopologues in the 1180–1210 cm$^{−1}$ region


Ibrahim Sadiek,[a*] Aleksandr A. Balashov,[b] Adrian Hjältén,[b] Michael Rey,[c] Oleg Egorov,[d] and Aleksandra Foltynowicz[b*]



Quantitative spectroscopic detection of dibromomethane, CH$_2$Br$_2$, for environmental monitoring, workplace safety, and exoplanetary studies is limited by the lack of accurate absorption cross-section data and rigorous spectroscopic models. We report the first high-resolution (6.3 MHz point spacing) absorption cross-section of CH$_2$Br$_2$ in the 1180 – 1210 cm$^{−1}$ region, measured using optical frequency comb Fourier transform spectroscopy. This spectral region is dominated by the strong CH$_2$ wagging ($\nu_8$) fundamental vibration, which is about 50 times stronger than the fundamental C−H stretch around 3077 cm$^{−1}$. The measurements resolve isotopologue-specific rovibrational features of CH$_2$$^{79}$Br$^{81}$Br, CH$_2$$^{79}$Br$_2$, and CH$_2$$^{81}$Br$_2$, and we assign rovibrational transitions of the $\nu_8$ fundamental and the overlapping $\nu_4+\nu_8-\nu_4$ hot bands using two methods. First, an empirical non-linear least square fit implemented in PGOPHER provides high-precision line assignment and spectroscopic constants, including accurate band origins, rotational constants, and quartic centrifugal distortion parameters, for the three isotopologues, covering rotational levels up to $K_a$ = 25 and $J$ = 144, with an average RMS residual of 0.00037 cm$^{−1}$ (11.1 MHz). Compared with previously reported band parameters retrieved from a fit to narrowband (1.78 cm$^{−1}$) supersonically cooled spectra (B. E. Brumfield *et al.*, *J. Mol. Spectrosc.*, 2011, **266**, 57-62), our fit provides much improved global agreement between measured and simulated spectra. In parallel, an ab initio-based effective Hamiltonian approach was used to model the complete rovibrational polyads, including weak hot-band transitions and polyad interactions inaccessible to purely empirical fits, and provided the first ab initio-based line intensities of CH$_2$Br$_2$ in the 8 μm spectral region.

## 1 Introduction

Dibromomethane, CH$_2$Br$_2$, is a naturally occurring halogenated volatile organic compound (HVOC) that is mainly emitted from the ocean.[1-4] Its spectroscopic monitoring is important for modelling the natural cycling of climate relevant traces gases, as CH$_2$Br$_2$ acts as a source of reactive halogen atoms that contribute to ozone depletion, and as a contributor to the stratospheric bromine budget due to the fast vertical transportation by the tropical deep convection.[3, 5] A potential anthropogenic source of brominated VOCs are treated ballast water systems,[6] where they are formed as disinfection by-products (DBPs) at concentrations substantially higher compared to their natural abundance.[7-9] Beyond terrestrial applications, methylated halogen species such as CH$_3$I, CH$_3$Br, CH$_3$Cl have recently been proposed as potential capstone biosignatures in exoplanetary atmospheres, owing to their comparatively low false-positive potential when used to confirm primary biosignatures such as O$_2$.[10] Dibromomethane may play a similar role, although this possibility remains unexplored.

Accurate spectroscopic models of CH$_2$Br$_2$ are therefore a prerequisite for (i) modelling the natural cycling of climate relevant traces gases, (ii) validating the existing models of the spread of ballast water discharge,[11] and assessing the ecotoxicological impact of DBPs on marine life and human health in harbour environments, and (iii) enabling the remote detection of halogenated biosignatures in Earth-like exoplanetary atmospheres.

Recent advances in optical frequency comb technology—combining high spectral resolution, broad spectral bandwidth, and absolute frequency accuracy—have enabled complete rovibrational band measurements with isotopologue specificity.[12-17] Moreover, the recent availability of fully stabilised comb sources and comb-based spectrometers operating in the 8 μm region[18-22] extends the reach of high-resolution, broadband precision spectroscopy into the long-wave mid-infrared (mid-IR) range.

Previously, we have reported the first frequency-comb-based high-resolution mid-IR spectra of CH$_2$Br$_2$ with isotopologue resolution in the region of 2960 to 3120 cm$^{−1}$,

a. *Experimental Physics V, Faculty of Physics and Astronomy, Ruhr University Bochum, 44780 Bochum, Germany. E-mail: ibrahim.sadiek@ruhr-uni-bochum.de*
b. *Department of Physics, Umeå University, 901 87 Umeå, Sweden. E-mail: aleksandra.foltynowicz@umu.se.*
c. *Laboratoire Interdisciplinaire Carnot de Bourgogne, UMR CNRS 6303, Université Bourgogne Europe, 9 Av. A. Savary, BP 47870, 21078 Dijon Cedex, France. E-Mail: michael.rey01@u-bourgogne.fr.*
d. *Laboratory of Theoretical Spectroscopy, V.E. Zuev Institute of Atmospheric Optics SB RAS, Tomsk 634055, Russia. E-Mail: oleg.egorov@iao.ru*

corresponding to the C–H stretching vibrations.[15] In that study, we revisited an earlier assignment of the asymmetric C–H vibration, $\nu_6$, around 3076 cm$^{-1}$ by Sadiek and Friedrichs,[23] which were based on continuous-wave cavity ring-down spectroscopy (CW-CRDS). We demonstrated that a cascade of hot-band transitions of the type $n\nu_4+\nu_6-n\nu_4$ (with $n \leq 3$) had been misassigned as isotopic shift of the three naturally abundant isotopologues: CH$_2$$^{79}$Br$^{81}$Br (mixed), CH$_2$$^{79}$Br$_2$ (light), and CH$_2$$^{81}$Br$_2$ (heavy).

In the long-wave mid-IR region, particularly around 1197 cm$^{-1}$ (or 8.35 um), CH$_2$Br$_2$ exhibits absorption features that are nearly 50 times stronger than those in the C–H stretching region. These absorptions arise from the CH$_2$ wagging vibration associated with the $\nu_8$ fundamental band and could offer a significantly enhanced sensitivity for spectroscopic detection, especially in workplace safety monitoring applications, and in astrophysics. Despite this advantage, no high-resolution absorption cross-section data are currently available for CH$_2$Br$_2$ in this spectral region. Previous room temperature spectra in the literature were acquired at low resolution using Fourier transform spectroscopy with incoherent light sources, which provided only overall band contours and lacked isotopic resolution and high-resolution cross-section data.[24] The only high-resolution study is a CW-CRDS measurements by Brumfield et al.,[25, 26] using a quantum cascade laser coupled to a supersonic expansion source. In that work, the assignment of the cold spectrum of the $\nu_8$ band was straightforward, with three resolvable rovibrational progressions attributed to the CH$_2$$^{79}$Br$^{81}$Br, CH$_2$$^{79}$Br$_2$, and CH$_2$$^{81}$Br$_2$ isotopologues. The CRDS measurements of Brumfield et al.,[25, 26] provided spectroscopic parameters that precisely reproduced the cold jet-cooled spectrum, however, their spectral coverage was limited to 1.78 cm$^{-1}$ around the band centre, which (i) restricted the fitted spectroscopic parameters to band origins, rotational constants, and a single quartic centrifugal constant, and (ii) limited the accuracy of these parameters for simulating room-temperature spectra. Consequently, reliable room-temperature cross-sections and comprehensive spectroscopic models for CH$_2$Br$_2$ in the $\nu_8$ region remain unavailable.

A major challenge in modelling the room-temperature spectra of brominated molecules is the existence of energetically low-lying vibrational states that retain significant thermal population. This leads to congested hot-band structure that overlaps and interferes with nearby fundamental transitions, similar to the C–H stretch region. However, such anticipated behaviour in the $\nu_8$ region has not yet been systematically investigated. Analogous hot-band progressions have been reported for other halomethanes, including iodomethane, CH$_3$I, where a single hot band near the fundamental C–H stretch bands has been observed,[27, 28] and diiodomethane, CH$_2$I$_2$, where up to 5 hot bands have been observed.[29] Another challenge is the nearly equal natural abundances of $^{79}$Br and $^{81}$Br, resulting in 2:1:1 population ratio of the mixed, light, and heavy isotopologues.

In this work, we present the first high-resolution room-temperature absorption cross-section of dibromomethane in the 1180 – 1210 cm$^{-1}$ spectral range, measured using a comb-based Fourier transform spectrometer (FTS) in the long-wave mid-IR infrared region.[21] These high-resolution spectra are used to develop a spectroscopic model that accounts for bromine isotopic abundances and the low-lying $\nu_4$ vibration. The empirical model, implemented in PGOPHER, includes the $\nu_8$ fundamental and the $\nu_4+\nu_8-\nu_4$ hot bands for the three isotopologues. This model is further benchmarked against ab initio line-list calculations based on a non-empirical effective Hamiltonian approach.[30] The combination of broadband, high-precision comb-based spectroscopy and the resulting global models provides a consistent and accurate description of the rovibrational structure of CH$_2$Br$_2$ in the 1180 – 1210 cm$^{-1}$ spectral range.

## 2 Experimental and computational details

### 2.1 Experimental Methods

The setup of the comb-based FTS experiment has been described previously,[21, 31, 32] and will only be briefly summarized here. It consists of a mid-IR frequency comb, a multi-pass absorption cell and an FTS. The mid-IR comb has a repetition rate, $f_{rep}$, of 125 MHz and spectral coverage of 1140 – 1280 cm$^{-1}$. It is produced by difference frequency generation (DFG) between pump and signal beams from the same Er:fiber oscillator[19] and is therefore inherently free from carrier-envelope-offset, $f_{ceo}$. The $f_{rep}$ was locked to the output of a tunable direct digital synthesizer locked to a GPS-referenced Rb clock.

The mid-IR beam passed through a Herriott multi-pass absorption cell (Thorlabs, HC10L/M-M02) with a path-length of 10.436(15) m. Analytical grade CH$_2$Br$_2$ sample (Acros organics – 99%) was contained in an Ace glass tube connected to the gas supply system and was allowed to evaporate into the cell under its vapour pressure. Prior to introducing CH$_2$Br$_2$ into the cell, the vacuum system was repeatedly purged with dry nitrogen gas and evacuated through a bypass valve to remove impurities. Afterwards the cell was filled to the desired pressure, measured using a pressure transducer (CERAVAC CTR 100 N 1 Torr range) with a manufacturer stated measurement uncertainty of 0.2%; the practical resolution was approximately 0.1 μbar. All measurements were performed at room temperature ranging between 23.1 and 24.2 °C.

After the absorption cell, the beam was coupled to a home-built fast-scanning FTS with a nominal resolution matched to the $f_{rep}$, and detected by two HgCdTe detectors in a balanced configuration. A beam of a narrow-linewidth CW diode laser with a wavelength of 1563 nm propagating on a path parallel to the comb beam in the FTS was used for calibrating the optical path difference (OPD).

We acquired two high-resolution spectra at 31 and 50.2 μbar using the sub-nominal resolution approach[33, 34]. We measured spectra at 20 $f_{rep}$ steps separated by 21 Hz, corresponding to 6.3 MHz point spacing in the optical domain. We scanned the $f_{rep}$ in alternating directions, and at each step we recorded a set of 20 interferograms. We repeated the scans 16 times, which gave a total of 320 interferograms at each $f_{rep}$ step and a total acquisition time of 7.5 h for all steps. A background spectrum was obtained from the average of 400 interferograms recorded at the first $f_{rep}$ step with the absorption cell evacuated. Half of the background interferograms were recorded before and half after the sample measurement. We normalized the averaged spectrum at each $f_{rep}$ step to the average background spectrum linearly interpolated to the wavenumber scale of the sample spectrum. We then corrected the remaining baseline by masking out the $CH_2Br_2$ band in the range from 1178 to 1212 cm$^{-1}$ and fitting a model of the baseline consisting of a 3$^{rd}$ order polynomial and a single sine function to account for an etalon appearing in the spectra. The baseline-corrected spectra measured at the different $f_{rep}$ steps were interleaved to produce a final spectrum with 6.3 MHz spectral point spacing.

Furthermore, to check the linearity of the dependence of integrated absorption on sample density, we acquired spectra at five different sample pressures between 10 and 50 μbar at a single $f_{rep}$ value (200 averages without interleaving). While these spectra are not fully resolved, the absorption integrated over the whole band can be compared across different pressures and to the same $f_{rep}$ step of the interleaved spectra. For more effective baseline correction in the spectra taken at one $f_{rep}$ step, we used a model of the $CH_2Br_2$ band based on the interleaved spectrum measured at 31 μbar. and fitted this band model together with the baseline model described above, instead of masking the $CH_2Br_2$ bands.

In the sub-nominal resolution method, the effective wavelength, $\lambda_{ref}$, of the reference laser used for OPD calibration is usually found by varying it in post processing to minimize instrumental line shape (ILS) distortions of individual lines.[34] This optimization was not possible on the $CH_2Br_2$ spectrum due to spectral congestion and scarcity of fully isolated absorption lines. However, weak $H_2O$ lines appeared away from the $CH_2Br_2$ bands, as a result of minor air leakage into the measurement cell during the measurement. Those $H_2O$ lines were used for optimizing $\lambda_{ref}$ by matching their positions in a preliminary measurement to a model simulated using parameters from the HITRAN2020 database.[35] We took the final $\lambda_{ref}$ value as the mean of the optima found for four different $H_2O$ lines located outside of $CH_2Br_2$ band. For all measurements reported here, $\lambda_{ref}$ was optimized by matching the $CH_2Br_2$ band contour to that in the optimized measurement. We estimate the frequency uncertainty of this procedure to be 2.4 MHz, which includes a 1.5 MHz uncertainty associated with line position determination arising from minimization of the instrumental line shape using water absorption lines, a ~1.8 MHz uncertainty in the HITRAN line centre positions of the reference $H_2O$ lines, and 0.2 MHz uncertainty from optimization of dibromomethane spectra shape to the reference measurement.

By comparing the spectra from the 16 $f_{rep}$ scans, we observed a linear decrease in $CH_2Br_2$ absorption of about 7% during the whole measurement. This decrease in absorption can be attributed to adsorption of the sample to the cell walls. We quantified the decrease by scaling a model of the $CH_2Br_2$ absorption to the spectrum at the first $f_{rep}$ step of each of the 16 $f_{rep}$ scans, together with the model of the baseline. The absorption model was obtained from the fully interleaved and averaged spectra measured at 31 μbar and 50.2 μbar, respectively. Since the absorption change was linear, we multiplied the absorption of the two measurements by half the percentual decrease, corresponding to 3.3% and 3.4% for the spectra at 31 μbar and 50.2 μbar respectively. The spectra of the linearity measurement were acquired in 15 minutes per pressure and thus negligibly affected by the decrease in sample density.

**2.2 Empirical PGOPHER simulations**

The spectral simulations and empirical rovibrational fitting were first performed using the PGOPHER software developed by C. M. Western.[36] This is an empirical effective Hamiltonian approach, in which spectroscopic parameters are optimised through nonlinear least-squares fitting to the experimentally assigned transitions.

Dibromomethane was treated as a near-prolate asymmetric top, with the Ir representation. The light and heavy isotopologues were treated within the $C_{2v}$ point group symmetry, whereas the mixed isotopologue was assigned $C_s$ symmetry owing to the reduced molecular symmetry introduced by isotopic substitution. Nuclear spin statistical weights of 9:7:7:9 were included for the $CH_2{}^{79}Br_2$ and $CH_2{}^{81}Br_2$ isotopologues, while nuclear spin statistics were not applied to the mixed isotopologue.

Fig. 1 illustrates the molecular structure of $CH_2Br_2$, including the principal axes of inertia and the symmetry planes of $C_{2v}$ isotopologues. The H–C–H plane contains the hydrogen atoms and carbon, reflecting the identical bromines, while the Br–C–Br plane contains both bromine atoms and carbon, reflecting the hydrogen atoms. The spectrum of $CH_2Br_2$ in the 1180 − 1210 cm$^{-1}$ range is dominated by the $CH_2$ wagging vibrations ($\nu_8$ band), which induces a change in the dipole moment along the *a*-axis, resulting in an *a*-type parallel band structure.

Ground-state rotational and centrifugal distortion constants were fixed to values obtained from microwave spectroscopy, taken from Niide *et al.*[37] for the mixed isotopologue and from Davis and Gerry[38] for the light and heavy isotopologues. Initial excited-state parameters for the $\nu_8$ fundamental band were taken from the narrowband cavity ring-down spectroscopy measurements of Brumfield *et al.*[26] and subsequently refined through fitting to the present broadband, room-temperature spectra. For the simulations of the $\nu_4+\nu_8-\nu_4$ hot bands, the band parameters of the low-lying lower state $\nu_4$ were taken from our previous simulations of the C–H stretch region,[15] and the upper state parameters were refined through fitting to the experimental spectrum. Overall, the empirical Hamiltonian model included the $\nu_8$ fundamental and the $\nu_4+\nu_8-\nu_4$ hot bands for all three isotopologues, resulting in a total of six vibrational bands. The PGOPHER input file of the simulations and assignment is given in the supporting information (S1).

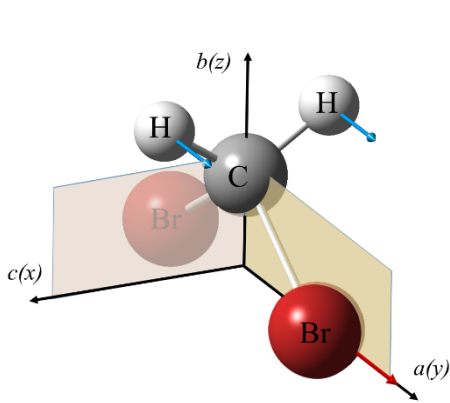

**Fig. 1** Molecular structure of $CH_2Br_2$. The blue and red arrows represent the displacement vectors and the direction of the transition dipole moment for the $\nu_8$ vibration. The principal axes of inertia and the two symmetry planes of the $C_{2v}$ isotopologues are also shown.

### 2.3 Non-empirical (*ab initio*) effective Hamiltonian simulations

While PGOPHER provides efficient, band-specific simulations by fitting a number of spectroscopic parameters to experimental transitions, its applicability is limited to the bands exclusively included in the model, and to the spectral coverage of the experimental spectrum used to empirically fit rovibrational levels. It cannot easily be extrapolated to unobserved vibrational states or very high rotational levels, and it does not generate a global line list. Furthermore, it provides only relative intensities. To overcome these limitations, we use the *ab initio* based effective Hamiltonian methodology, proposed in Ref.[30] as a complementary approach. This method combines the simplicity of the traditional effective Hamiltonian approach, such as PGOPHER, with the completeness of pure variational calculations to provide globally accurate model for the construction of comprehensive line lists.

Following our previous *ab initio* works, in particular, ref.[39] and [40], the potential energy surface (PES) and dipole moment surface (DMS) of $CH_2Br_2$ were both expressed as a polynomial expansion in terms of irreducible tensor operators (ITOs) built from internal coordinates and adapted to the $C_{2v}$ and $C_s$ point groups. Both the PES and the DMS were then expanded as Taylor series in normal-mode coordinates up to sixth order, while the Eckart–Watson kinetic energy operator[41], implemented in the TENSOR code[42-45], was employed. In order to determine the expansion coefficients, *ab initio* electronic structure calculations were performed on grids of reference nuclear configurations.

In the case of the PES, the RHF-CCSD(T)-F12b method implemented in the MOLPRO package[46, 47] was employed in the first step in combination with the VQZ-PP-F12 and CVQZ-F12 basis sets for bromine, and for hydrogen and carbon, respectively. In the second step, corrections were included for: (i) scalar relativistic effects (within the Cowan-Griffin approach); (ii) the diagonal Born Oppenheimer correction (computed at the RHF-CCSD/cc-pVDZ level using the CFOUR package[48, 49]); and (iii) high-order electron correlation effects (evaluated using the CCSDT(Q)/VDZ and CCSDTQ/VDZ methods within the MRCC package[50, 51]). The resulting *ab initio* energies for the 5368 nuclear configurations were fitted using ITOs up to the sixth order, yielding 460 expansion coefficients. The root-mean-square deviation of the fit was 0.163 cm$^{-1}$.

The *ab initio* calculations of the electric dipole moment were computed using the finite difference method as the first derivative of the potential energy of nuclei with respect to the electric field. The RHF-CCSD(T)-F12b/AVQZ level of theory, as implemented in MOLPRO, was employed to solve the electronic problem. The number of nuclear configurations was 5368 for the $A_1$ component, and 4444 for the $B_1$ and $B_2$ components. The RMS errors of the fit to the refence *ab initio* dipole moment values did not exceed $3.6 \times 10^{-5}$ Debye using 418, 351 and 397 expansion coefficients for the $A_1$, $B_1$, and $B_2$ components of the DMS, respectively.

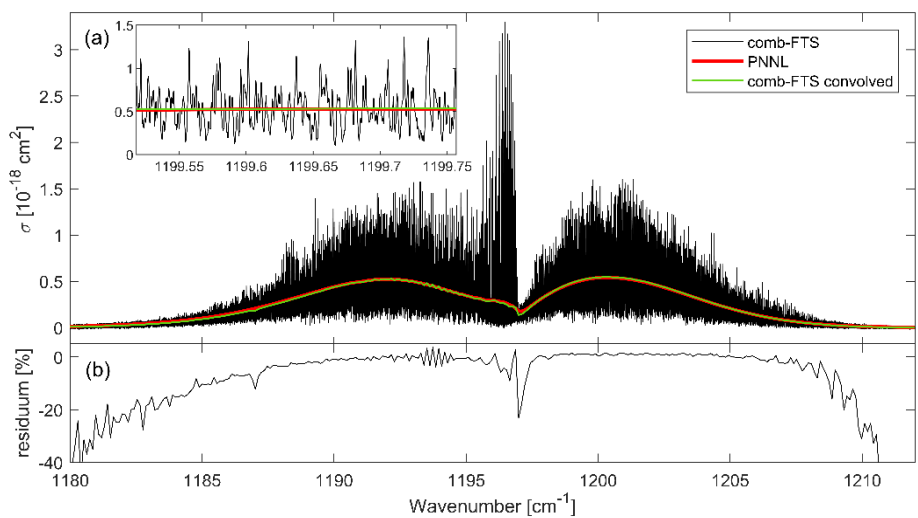

**Fig. 2** (a) Absorption cross-section of $CH_2Br_2$ measured in this work at 31 µbar using the comb-based FTS (black) compared to the PNNL data[24] (red). The green trace shows the cross section from this work convolved with the *sinc* function to mimic the resolution of PNNL spectra. The inset shows a zoom around 1199.65 cm$^{-1}$. (b) The relative discrepancy between the convolved comb-FTS measurement and the PNNL data.

The global effective approach begins with a full variational nuclear-motion Hamiltonian $H^{(J,C)}$ computed for a given symmetry block, defined by rotational quantum number $J$ and symmetry $C$, in a primitive vibrational basis $|\gamma, J, C\rangle$, where $\gamma$ represents all other quantum numbers. This Hamiltonian includes all vibrational couplings and rotational interactions within the block, capturing the complete rovibrational structure. A unitary transformation $T^{(J,C)}$ is then applied to the Hamiltonian to obtain a block-diagonal effective Hamiltonian $H^{(J,C,P)}$, truncated to a chosen polyad $P$. This block-diagonalization ensures that strongly interacting states within a polyad are treated exactly, while weakly interacting states outside the polyad are effectively decoupled. The transformation also produces a set of effective eigenvectors $\widetilde{U}(J,C)$ and eigenvalues $E_{vr}$ corresponding to vibrational-rotational energy levels. In this work, the eigenpairs were computed up to $J$ = 20. The variational energies $E_{vr}$ can be replaced with experimental energies $E_{vr}^{obs}$, allowing refinement of the effective Hamiltonian with minimal computational effort. In case of $CH_2Br_2$, the empirically determined energy levels from PGOPHER simulations were used as $E_{vr}^{obs}$ to refine the Hamiltonian. To this end, a small subset of so-called diagonal parameters was refined for the ground state and for the $\nu_4$, $\nu_8$ and $\nu_4 + \nu_8$ vibrational bands. Here, 'diagonal' denotes that the powers of the creation and annihilation operators are equal.

The same transformation is applied to the dipole moment operators to compute line intensities to ensure that predicted transition strengths are consistent with the effective Hamiltonian energy levels. This approach allows rapid calculation of high-$J$ rotational levels while retaining global accuracy, and has been successfully applied to molecules such as $CH_4$[52], and more recently for $SiF_4$[39] and $CHF_3$.[40] However, $CH_2Br_2$ presents a more challenging case as highlighted earlier for several reasons: (i) As a heavy halogenated asymmetric top, it has a much denser rotational structure and more closely spaced energy levels than the relatively light and symmetric molecules mentioned above. (ii) The presence of two bromine atoms with the nearly equal natural isotopic abundance of $^{79}Br$ and $^{81}Br$ introduces multiple isotopologues with different symmetries. Note that the $C_s$ isotopologue $^{12}CH_2{}^{79}Br^{81}Br$ is the most abundant species. (iii) The coexistence of low-lying vibrational state which increase the density of state at room temperature, and hence requires careful treatment of rovibrational couplings. Such low-lying vibrational states result in a substantial number of hot bands, similar to $CH_4$ at high temperatures (>1000 K).

A more complete description of the *ab initio* based effective Hamiltonian model for $CH_2Br_2$ isotopologues will be published in a separate paper.

## 3 Results and discussion

### 3.1 High-resolution absorption cross-section

Fig. 2(a) presents the high-resolution absorption cross-section of $CH_2Br_2$ in the region from 1180 $cm^{-1}$ to 1212 $cm^{-1}$ retrieved from the spectrum measured at 31 µbar with a point spacing of 6.3 MHz (black), compared to the cross-section from the Pacific Northwest National Laboratory (PNNL) database (red),[24] which has a resolution of 3.4 GHz. The inset shows a zoomed spectral window around 1199.65 $cm^{-1}$ demonstrating the dramatic improvement in spectral resolution compared to the Fourier transform infrared (FT-IR) spectrum from PNNL, which does not resolve the rotational structures at all. As a comparison the frequency comb data is also plotted convolved with a *sinc* function with a zero-crossing spacing of 3.4 GHz (green) to mimic the resolution of the PNNL spectra. Fig. 2(b) shows the relative difference in cross-section between the convolved spectrum and the PNNL spectrum. The comb spectrum agrees with PNNL to within a few percent throughout most of the $CH_2Br_2$ bands, deviating significantly only on the wings where the absorption is small and slight differences in baseline between the two measurements would have a large impact on the relative discrepancy. The experimental cross-section is provided in the supporting information (S2).

Figure 3 shows the result of a measurement of pressure dependence of integrated absorption linearity performed to evaluate the uncertainty of absorption cross-section. For linearity measurements, we acquired spectra at five different pressures. The sample cell was evacuated between each measurement. The absorption coefficient integrated from 1178 $cm^{-1}$ to 1212 $cm^{-1}$ excluding the saturated Q-branch at 1196-1197 $cm^{-1}$ is shown as black dots. The inset of Fig. 3 shows a $CH_2Br_2$ spectrum with spectral range used for absorption coefficient integration highlighted in black. The dotted line shows a linear fit to the linearity measurement data. The fit included an intercept as a free parameter to evaluate any potential offset in the pressure measurement, but it was not statistically significant. The plot also includes the integrated absorption coefficient of the two interleaved measurements (red crosses) where the data point at 31 µbar corresponds to the spectrum shown in Fig. 2. The spectrum at 50.2 µbar was required to enable the assignments of weak transitions of hot bands of light and mixed isotopologues.

The standard deviation of all data points in lower panel of Fig. 3 is 2%, which is comparable to the difference of 2.7% in the cross-sections of the comb-FTS and PNNL spectra integrated over the same range as for the linearity measurement. However, despite this very good agreement, comparison of the integrated absorption of the spectra in the linearity series — using either the band shape model described in Section 2.1 or masking the band region during baseline correction —indicates that baseline correction can contribute up to 10% uncertainty to the slope in Fig. 2 and to the cross-section in Fig. 1. The band shape model is used for the results presented here, as it allows a simultaneous fit of both the baseline and absorption feature across the full spectral range, whereas masking the band region constrains the baseline fit only to the surrounding spectral regions and therefore increases its sensitivity to chosen etalons and polynomial order.

Beyond the comparison with PNNL, we estimated a minimum detectable absorption coefficient from the standard deviation of the noise in the baseline of the spectrum at 31 µbar to be $\alpha_{min}$ of 4.6 × $10^{-6}$ cm$^{-1}$ in measurement time of $t$ = 7.5 h. This corresponds to a noise equivalent absorption sensitivity, NEAS = $\alpha_{min} \cdot t^{1/2}$ = 7.5 × 10$^{-4}$ cm$^{-1}$ Hz$^{-1/2}$. Taking into account the spectral coverage of the comb with $M$ = 2.4 × 10$^5$ spectral elements in the interleaved spectra, we also report the figure of merit FoM = NEAS·$M^{-1/2}$ = 1.5 × 10$^{-6}$ cm$^{-1}$ Hz$^{-1/2}$ per spectral element.

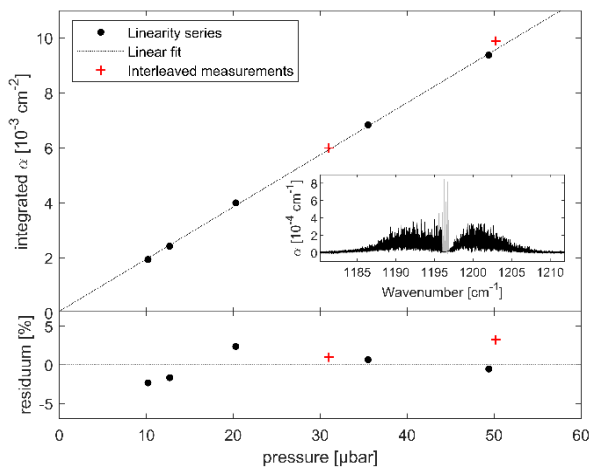

**Fig. 3** The absorption coefficient integrated from 1178 cm$^{-1}$ to 1212 cm$^{-1}$ excluding the Q-branch (see black trace in the inset where grey data points were excluded) in the linearity series (black dots) and the two interleaved measurements (red crosses). The dotted line shows a linear fit to the linearity series. The bottom panel shows the fit residuum expressed as a percentage of the absorption.

## 3.2 Spectral simulations and analysis

In this section, the rovibrational spectra of CH$_2$Br$_2$ in the 1180–1220 cm$^{-1}$ range are analysed for all three isotopologues: CH$_2^{79}$Br$^{81}$Br, CH$_2^{79}$Br$_2$, and CH$_2^{81}$Br$_2$. The analysis combines empirical effective Hamiltonian simulations implemented in PGOPHER with benchmark comparisons to non-empirical *ab initio*-based effective Hamiltonian line lists.

### 3.2.1 Empirical effective Hamiltonian simulations

The parallel nature of the $\nu_8$ band might suggest a straightforward assignment, however, as mentioned earlier the presence of the low-lying $\nu_4$ vibration (~172 cm$^{-1}$) and the nearly equal abundance of the two bromine isotopes significantly complicates the spectra at room temperature. For an equilibrated sample, the first hot band ($\nu_4+\nu_8-\nu_4$) is predicted to appear near the fundamental band with a population ratio of 0.44, based on the Maxwell-Boltzmann distribution. Consequently, each isotopologue exhibits both cold and hot parallel bands, all of which must be simulated to reproduce the room-temperature spectra of CH$_2$Br$_2$ with naturally abundant $^{79}$Br and $^{81}$Br.

Figure 4 shows the measured high-resolution spectrum at 31 μbar of CH$_2$Br$_2$ together with simulations of the fundamental $\nu_8$ and the hot $\nu_4-\nu_8+\nu_4$ bands of each isotopologue: CH$_2^{79}$Br$^{81}$Br (green), CH$_2^{79}$Br$_2$ (red), and CH$_2^{81}$Br$_2$ (blue). The simulations of each isotopologue are offset for clarity, and for each isotopologue the hot and the fundamental band simulations are also offset relative to one another. The strong rovibrational features at the band centre of the measured spectra are saturated due to the strong absorption of CH$_2$Br$_2$. The assignment was enabled through the measurements at lower pressure of 31 μbar for the strong fundamental band transitions, and at a higher pressure of 50.2 μbar for the weak transitions of the light and heavy isotopologues and for the hot bands. The predicted 2:1:1 intensity ratio of CH$_2^{79}$Br$^{81}$Br, CH$_2^{79}$Br$_2$, and CH$_2^{81}$Br$_2$ isotopologues, explained by the isotope abundance, can be seen in Fig. 4. In addition, the intensities of the hot bands roughly match the predictions from Maxwell-Boltzmann statistics for each isotopologues. It should be noted again that PGOPHER provides only relative intensities, particularly for hot bands, unless a proper partition function is specified; absolute line strengths or absorption cross-sections require additional scaling and proper treatment of coupling between vibrational states. These limitations can be addressed in the complementary *ab initio*-based non-empirical simulations (vide infra).

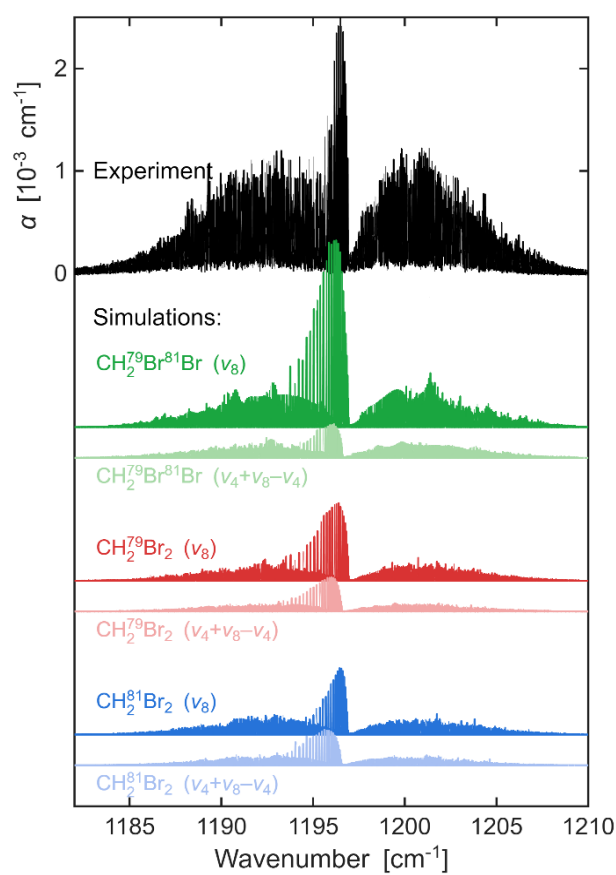

**Fig. 4** High resolution absorption coefficient ($\alpha$, in cm$^{-1}$) measured at 31 μbar using the comb-based FTS together with the PGOPHER simulations of the fundamental $\nu_8$ and the $\nu_4+\nu_8-\nu_4$ hot bands of CH$_2^{79}$Br$^{81}$Br (green), CH$_2^{79}$Br$_2$ (red), and CH$_2^{81}$Br$_2$ (blue). The overall simulations involve the sum of all six vibrational bands.

Initial simulation of the $\nu_8$ spectra utilized band origins and rotational constants from the CRDS measurements of Brumfield *et al.*,[26] which were obtained from a fit to a narrow spectral coverage of 1.78 cm$^{-1}$ around the band centre, measured in the jet-cooled sample. These parameters provide a good initial match with the Q-subcluster rovibrational features near the band centres of the measured comb spectra at room temperature. However, the agreement deteriorated far from the band centre, and the parameters were lacking for the hot band transitions.

**Band parameters from line fit.** The first set of assignments involved the progression of the strong, sharp spectral features at the centre of the $\nu_8$ bands for all the three isotopologues, shown in Fig. 4. For each isotopologue, these features represent a series of Q-subcluster progressions characteristic of a-type parallel vibrations. Each single feature is composed of a series of tightly packed transitions sharing the same $K_a$ for the lower and upper states: $Q(J'', K_a'', K_c'')$: $(\nu', J' = J'', K_a' = K_a'', K_c' = K_c'' \pm 1) \leftarrow (\nu'', J'', K_a'', K_c'')$. The assignment of these Q-subclusters was possible up to $K_a$ values of $22-25$, thanks to the broad spectral coverage of the frequency comb. This broad coverage of the full band directly impacts the accuracy of the obtained band parameters and the 'global' match of the experimental spectra to the simulations (vide infra).

After fitting the assigned Q-subclusters, partially resolved transitions in the $P$ and $R$ branches could be assigned. Similarly, for the $\nu_4+\nu_8-\nu_4$ hot bands, the assignment started with the Q-subclusters, first for the mixed isotopologue and subsequently for the light and heavy isotopologues. For these bands, the fit was mostly limited to the Q-subcluster progressions, as unambiguous assignments of P- and R- branch transitions was hindered by the cross-interfering absorption of fundamental band transitions. It should be noted that higher-order hot bands, $2\nu_4+\nu_8-2\nu_4$ and $3\nu_4+\nu_8-3\nu_4$, are predicted to reach population ratios of ~0.19 and ~0.08 at room temperature, respectively. However, assignment of their Q-subcluster progressions was challenging due to the overlap with other strong bands. Overall, 6298 transitions were assigned among all simulated bands. These assigned transitions were used in non-linear least square fit to provide effective band parameters for the $\nu_8$ and $\nu_4+\nu_8-\nu_4$ bands of the three isotopologues. In addition, they were used to refine the effective Hamiltonian of the non-empirical ab initio-based calculations as described earlier.

Figure 5 (a) shows the overall simulated spectrum (inverted for clarity) after fitting the assigned transitions, together with the experimental spectrum measured at $CH_2Br_2$ pressure of 50.2 μbar. Red vertical bars at the top of panel (a) indicate the 297 transitions assigned by Brumfield et al.[26] within the 1.78 cm$^{-1}$ measurements window. As shown in this figure, an overall very good match in the band structure between the experiment and simulation is observed. The lower panels present zoomed-in spectral windows highlighting the quality of the match near the band centre and farther from it. For example, at the band centre [panel (e)], the Q-subcluster progressions for both the $\nu_8$ and $\nu_4+\nu_8-\nu_4$ bands show excellent agreement with the experiment for the three isotopologues, as indicated by the faithful reproduction of all strong and weak absorption peaks. Close to the band centre, but apart from these Q-subclusters [panel (d)], still a very good match is obtained. Even farther from the band centre, panels (b) and (g), several sharp rovibrational transitions could be identified.

Table 1 summarizes the fitted band parameters of the fundamental $\nu_8$ and the hot $\nu_4+\nu_8-\nu_4$ bands for the $CH_2^{79}Br^{81}Br$

isotopologue, together with the $\nu_8$ band parameters reported by Brumfield et al.[26] For the $\nu_8$ band, the band origin and rotational constants obtained in the present work agree with those of Brumfield et al. within ≈1σ of the combined uncertainty, with the band origins differing by only $2.1 \times 10^{-5}$ (~0.6 MHz). The uncertainty of the $\nu_8$ band origin in the present work is about four times smaller than that reported by Brumfield et al., while the uncertainties of the rotational constants are lower by more than one order of magnitude. In addition, higher order centrifugal distortion constants could be floated in the present fit, while in the work of Brumfield et al. these parameters were fixed to their ground state values. Also listed in this table are the number of assigned $K_a$ values, $N(K_a)$, the maximum assigned rotational levels, $J_{max}$, and the total number of transitions. Since dibromomethane is a near-prolate asymmetric top with a Ray's asymmetry parameter of approximately −0.996, the rotational energies depend strongly on $K_a$ but only weakly on $K_c$, leading to very similar frequencies for many distinct rovibrational transitions that differ only in $K_c$. Several transitions satisfying $E(J, K_a, K_c \pm 1) \approx E(J, K_a, K_c)$ may become coincident within the Doppler width of the profiles and experimentally unresolved. These coincident transitions are not restricted to the Q-subclusters near the band centre but also occur for several transitions in the P and R branches. In the present analysis, these coincident transitions, together with the Q-subclusters, are treated as single effective transitions in the fitting procedure. Table 1 additionally reports the number of unique frequencies assigned, $N_{uniq}$, corresponding to the number of experimentally distinguishable line positions used in the fit.

Table 1 The band parameters of the fundamental $\nu_8$ and the $\nu_4+\nu_8-\nu_4$ hot band for the $CH_2^{79}Br^{81}Br$ isotopologue (all in cm$^{-1}$), number of $K_a$ and $J_{max}$ assignments, the number of assigned transitions, N(trans), number of unique frequencies assigned to the experiment $N_{uniq}$, and their root-mean-square (RMS) values (in cm$^{-1}$). Values in parentheses are 1σ uncertainties.

|  | $\nu_0$ | $\nu_8$ | | $\nu_4+\nu_8$ |
|---|---|---|---|---|
|  | Niide et al.[37] | This work | Brumfield et al.[26, a] | This work [a] |
| Origin |  | 1196.957031(14) | 1196.957052(52) | 1367.727286(37) |
| A | 0.8675192 | 0.86265314(19) | 0.8626518(25) | 0.86629162(46) |
| B | 0.0408047 | 0.04082189(2) | 0.0408228(16) | 0.04076770(20) |
| C | 0.0392537 | 0.03923653(2) | 0.0392382(14) | 0.03924246(20) |
| $\Delta D_K \times 10^{7\ b}$ |  | −1.956(5) | −2.17(22) | −2.064(9) |
| $D_{JK} \times 10^7$ | −3.810 | −3.813(2) |  |  |
| $D_J \times 10^9$ | 7.749 | 7.8032(3) |  |  |
| $d_1 \times 10^{10}$ | −6.44 | -6.531(7) |  |  |
| $d_2 \times 10^9$ | −1.03 | −1.0507(4) |  |  |
|  |  |  |  |  |
| $N(K_a)$ |  | 24 | 12 | 25 |
| $J_{max}$ |  | 144 | 13 | 80 |
| N(trans) |  | 3549 | 123 | 386 |
| $N_{uniq}$ |  | 1512 | 86 | 97 |
| RMS |  | 0.00036 | 0.00023 | 0.00043 |

[a]: Values of $D_{JK}$, $D_J$, $d_1$, and $d_2$ were fixed to their ground state constants based on microwave measurements of Niide et al.[37]
[b]: $\Delta D_K = D_K' - D_K''$, with $D_K''(\nu_8) = 1.290 \times 10^{-5}$ cm$^{-1}$ and $D_K''(\nu_4) = 1.240 \times 10^{-5}$ cm$^{-1}$.

The root-mean-square (RMS) error of the fit is defined as:[36]

$$\text{RMS} = \sqrt{\frac{\sum_i^{n_{obs}}(\text{Obs}_i - \text{Calc}_i)^2}{n_{obs} - n_{par}}}$$

where $n_{obs}$ is the total number of assigned transitions, $Obs_i$ and $Calc_i$ are the observed and calculated frequencies of the i$^{th}$ transition, respectively, and $n_{par}$ is the number of floated parameters in the least squares fitting.

For the $\nu_8$ fundamental band of $CH_2^{79}Br^{81}Br$, an RMS value of 0.00036 cm$^{-1}$ (or 10.8 MHz) was obtained. This value is ~1.6 larger than the value of 0.00023 cm$^{-1}$ reported by Brumfield et al.[26] However, the present spectroscopic model provides a more global description of the rovibrational structure, as it includes assignments up to $K_a$ = 24 and $J_{max}$ = 144, compared to $K_a$ = 12 and $J_{max}$ = 13 in the earlier study.[26] Actually, the RMS value reported by Brumfield et al. should be interpreted with consideration of the $N_{uniq}$ contributing to the fit. While Q-subclusters were treated as single observed frequencies in their analysis, several coincident pairs of P- and R-branch transitions with identical $J$ and $K_a$ and $\Delta K_c=\pm 1$ were counted as separate observed transitions, despite corresponding to the same experimentally unresolved line position. As discussed above, such coincidences arise naturally from the near-prolate asymmetric-top structure of $CH_2Br_2$. When the RMS is evaluated using the $N_{uniq}$ rather than the total number of assigned transitions, their RMS values would increase by ~20% on average for all isotopologues. It should also be noted that at room temperature the Q-subclusters assigned in the present work consist of substantially larger number of tightly packed transitions compared to those observed under supersonic jet conditions (with temperatures of ~20 K).[26] As a result, the transitions contributing to each cluster span a broader Obs–Calc distribution at room temperature than the few transitions forming each cluster at low temperature (vide infra). Consequently, RMS values should be interpreted in the context of spectral congestion, and the extent of rovibrational state coverage.

Finaly, the RMS value of 10.8 MHz for the $\nu_8$ band obtained here corresponds to only ~1/3 of the Doppler FWHM of $CH_2Br_2$ at room temperature (33 MHz), demonstrating the high precision of comb-based FTS even at ambient conditions. By comparison, the RMS of ~8 MHz reported for the jet-cooled cw-CRDS spectrum corresponds to roughly 60% of the linewidth (of 13.5 MHz) of the narrow transitions measured in the jet. These comparisons further illustrate that, despite the broader Doppler-limited conditions and substantially larger rovibrational state coverage considered in the present work, the achieved fit precision remains competitive with that obtained in jet-cooled measurements. Combining such broadband comb measurements with supersonic sources or buffer gas cooling methods,[20, 22] will provide an unambiguous assignment with ultimate accuracy and precision.

For the $\nu_4+\nu_8-\nu_4$ hot band, an RMS value of 0.00043 cm$^{-1}$ (or 12.9 MHz) was obtained from 97 unique frequencies, with assignments extending up to $K_a$ = 25 and $J_{max}$ = 80 (see Table 1). The fit involved five floated parameters ($n_{par}$ = 5), limited to the band origin, the rotational constants, and the quartic centrifugal constant $D_K$, while the remaining parameters listed in Table 1 were fixed to their $\nu_0$ values in both lower and upper states.

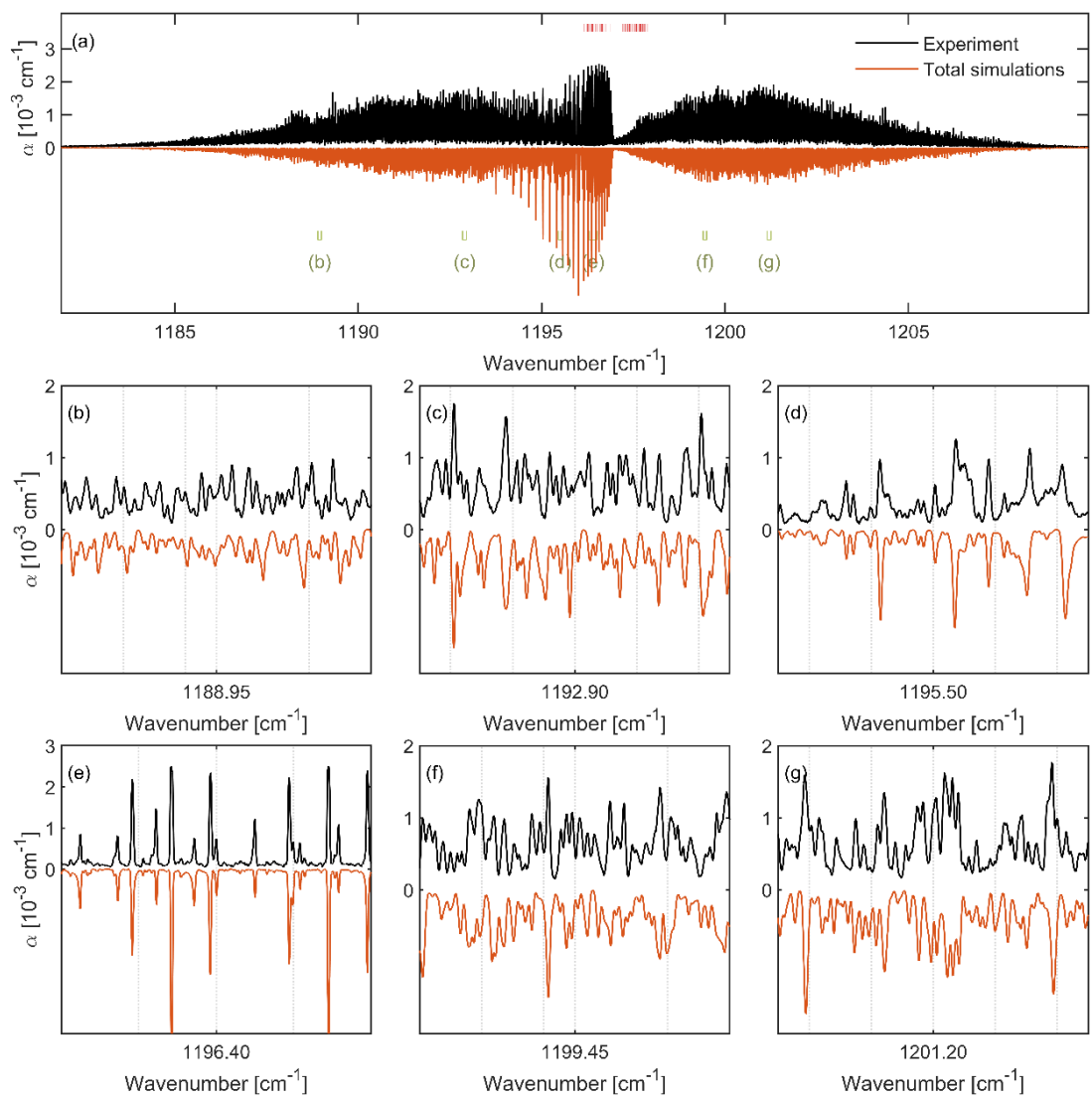

**Fig. 5** (a) Comparison of the absorption coefficient measured at 50.2 μbar (black) of $CH_2Br_2$ with the overall PGOPGHER simulations (orange). The central part of the fundamental band is saturated because of the strong absorption signal at this relatively high pressure. The red vertical bars in the top panel indicate the assigned 297 transitions (or $N_{uniq}$ = 213 unique frequencies) for the $\nu_8$ band of the three isotoplogues by Brumfield *et al.* [26] from CRDS measurements combined with a supersonic jet. (b) – (g) Zoomed-in spectral windows at different parts of the spectrum.

Table 2 The band parameters of the fundamental $\nu_8$ and the hot $\nu_4+\nu_8-\nu_4$ bands for the CH$_2$$^{79}$Br$_2$ and CH$_2$$^{79}$Br$_2$ isotopologues (all in cm$^{-1}$), the number of $K_a$ and $J_{max}$ assignments, the number of assigned transitions, N(trans), the number of unique frequencies assigned to the experiment (N$_{uniq}$) and their root-mean-square (RMS) values (in cm$^{-1}$). Values in parentheses are 1σ uncertainties.

|  | CH$_2$$^{79}$Br$_2$ | | | | CH$_2$$^{81}$Br$_2$ | | | |
|---|---|---|---|---|---|---|---|---|
|  | $\nu_0$ | $\nu_8$ | | $\nu_4+\nu_8$ | $\nu_0$ | $\nu_8$ | | $\nu_4+\nu_8$ |
|  | Davis & Gerry[38] | This work [a] | Brumfield et al.[26, a] | This work [a] | Davis & Gerry[38] | This work | Brumfield et al.[26, a] | This work [a] |
| Origin |  | 1196.982451(22) | 1196.982565(56) | 1368.679265(46) |  | 1196.931404(21) | 1196.931350(46) | 1366.713846(45) |
| A | 0.8683441 | 0.86347446(33) | 0.8634374(31) | 0.86696106(47) | 0.8667564 | 0.86189101(29) | 0.8618897(32) | 0.86545019(52) |
| B | 0.0413131 | 0.04132770(8) | 0.0413299(24) | 0.04129157(90) | 0.0402973 | 0.040306916(80) | 0.0403157(14) | 0.03997084(68) |
| C | 0.0397255 | 0.03971131(7) | 0.0397103(22) | 0.03973530(97) | 0.0387823 | 0.038772105(82) | 0.0387663(13) | 0.03845840(79) |
| $\Delta_K \times 10^5$ | 1.2922 | 1.27231(8) | 1.2676(22) | 1.24040(8) | 1.2879 | 1.26860(6) | 1.2667(33) | 1.2071(1) |
| $\Delta_{JK} \times 10^7$ | −3.8403 | −3.8388(7) |  |  | −3.7483 | -3.8186(6) |  |  |
| $\Delta_J \times 10^9$ | 7.9321 | 8.139(2) |  |  | 7.5662 | 7.596(2) |  |  |
| $\delta_J \times 10^{10}$ | 5.2279 | 4.699(56) |  |  | 4.8773 | 6.06(8) |  |  |
| $\delta_K \times 10^8$ | 3.93 | 3.9247(2) |  |  | 3.782 | 3.7726(4) |  |  |
|  |  |  |  |  |  |  |  |  |
| N($K_a$) |  | 25 | 13 | 25 |  | 24 | 11 | 22 |
| $J_{max}$ |  | 71 | 13 | 42 |  | 101 | 12 | 31 |
| N(trans) |  | 663 | 92 | 254 |  | 1257 | 82 | 189 |
| N$_{uniq}$ |  | 265 | 67 | 42 |  | 575 | 60 | 24 |
| RMS |  | 0.00037 | 0.00027 | 0.00040 |  | 0.00037 | 0.00022 | 0.00028 |

[a]: Values of $\Delta_{JK}$, $\Delta_J$, $\delta_J$, and $\delta_K$ were fixed to their ground state constants based on microwave measurements of Davis and Gerry.[38]

Table 2 summarizes the band parameters of the fundamental $\nu_8$ band for the light and heavy isotopologues obtained from the present room-temperature comb measurements and those reported by Brumfield et. al.[26] from CRDS at the supersonic temperature. Also included in this table are the band parameters of the $\nu_4+\nu_8-\nu_4$ hot bands obtained from the present work. Similar to the mixed isotopologue, the uncertainties of the $\nu_8$ band origins of the light and heavy isotopologues obtained here are about a factor of two smaller than those reported by Brumfield et al.,[26] while for the rotational constants they are lower by about one order of magnitude. For CH$_2$$^{79}$Br$_2$, the difference in the $\nu_8$ band origin between the two studies is 1.14 × 10$^{-4}$ cm$^{-1}$, which is within ~2σ of the combined uncertainties. The rotational constants A, B, and C differ by 3.7 × 10$^{-5}$ cm$^{-1}$, −2.2 × 10$^{-6}$ cm$^{-1}$, and 1.0 × 10$^{-6}$ cm$^{-1}$, corresponding to ≈80σ, 9σ, and 5σ of the combined uncertainties, respectively, and have a very large impact on the residuals of the assigned high J transitions (see below). Similarly, for the $\nu_8$ band of CH$_2$$^{81}$Br$_2$, the band origin obtained in the two studies differ by only 5.4 × 10$^{-5}$ cm$^{-1}$ (~1.6 MHz), corresponding to about 1σ of the combined uncertainty. In contrast, the rotational constant A, B, and C differ by several σ, and again this difference is significant, and it reflects the impact of extended J and $K_a$ coverage of the present assignment.

By comparing the RMS values for the $\nu_8$ band of the light and heavy isotopologues with those from Brumfield et al.,[26] the same arguments as for the mixed isotopologue apply. After accounting for the number of unique frequencies used in the fits, the RMS values obtained here are approximately 1.4 times larger for both CH$_2$$^{79}$Br$_2$ and CH$_2$$^{81}$Br$_2$ than those reported by Brumfield et al. Importantly, the present results are based on substantially broader rovibrational coverage, extending up to $K_a$ = 25 and $J_{max}$ = 71 for CH$_2$$^{79}$Br$_2$ (compared to $K_a$ = 13 and $J_{max}$ = 13), and up to $K_a$ = 24 and $J_{max}$ = 101 for CH$_2$$^{81}$Br$_2$ (compared to $K_a$ = 11 and 12). As a result, the derived parameters represent global effective values that provide an improved description of the rovibrational structure, as reflected in the fit residuals.

**Fit residuals.** Fig. 6 shows the residuals of the least squares fit to wavenumbers of a total of 3034 uniquely assigned transitions in the $\nu_8$ band as a function of J and $K_a$ quantum numbers for the CH$_2$$^{79}$Br$^{81}$Br (green), CH$_2$$^{79}$Br$_2$ (red), and CH$_2$$^{81}$Br$_2$ (blue) [panel (a)]. The residuals are randomly scattered around zero for transitions with quantum numbers up to J = $J_{max}$ and $K_a$ = N($K_a$) for each isotopologue – see Tables 1 and 2. As can be seen in Fig. 6, the assignment of clustered Q(J, $K_a$, $K_c$) transitions to a single observed peak results in larger residual spreads for these transitions. For example, at the highest $K_a$ = 24 of CH$_2$$^{81}$Br$_2$ in Fig. 6(a), this large vertical spread of residuals comes from the multiple Q(24) sub-cluster transitions being assigned to the same frequency. Such spread adds to the overall RMS error budget.

Applying the band parameters reported by Brumfield et al.,[26] to the broadband dataset assigned here results in systematic deviations at higher J values, as shown in panel (b) (note the different vertical scale). In contrast, simulations based on the present parameters exhibit significantly reduced deviations across the full range of J and $K_a$. The large residuals observed at high J and $K_a$ when using band parameters from earlier work are attributed to small difference in rotational constants (A, B, and C) and to the use of ground state centrifugal distortion constants ($D_J$, $D_{JK}$, $D_K$) for the excited state, as the limited number of transitions did not allow for floating theses parameters. Such high sensitivity of the fit residuals to small differences in spectroscopic parameters is enabled by the broadband coverage and high frequency precision of the comb-based Fourier transform spectrometer.

The fit residuals of the $\nu_4+\nu_8-\nu_4$ hot band transitions for all three isotopologues are shown in panel (c) of Fig. 6. No systematic trends are observed as a function of $J$ or $K_a$, although systematic deviations may appear at $J > J_{max}$ assigned here.

Finally, it should be noted that the average (from all isotopologues) RMS error of ~0.00037 cm$^{-1}$ (~11.1 MHz) is about 4 times larger than the experimental frequency uncertainty (2.4 MHz – see Experimental Section). This clearly indicates that the average error is limited by the spectroscopic model rather than the experimental precision. This is not unexpected given the complexity of the room-temperature spectra and the existence of several other additional unassigned absorption peaks, most likely due to higher-order hot bands (e.g., $2\nu_4+\nu_8-2\nu_4$ and $3\nu_4+\nu_8-3\nu_4$). In addition, the natural isotopic composition of bromine leads to overlapping contributions from multiple isotopologues, further increasing the spectral congestion. These challenges limit the band-by-band empirical analysis and motivate the use of a more global spectroscopic description capable of accounting for interactions between multiple vibrational states simultaneously.

### 3.2.2 Non-empirical (*ab initio*-based) spectral calculations

In contrast to the band-by-band analysis performed with PGOPHER, the simulations based on the *ab initio*-based effective Hamiltonian formalism[30] follow a more global description of the rovibrational structure. In this approach, vibrational states that are connected through anharmonic resonances are grouped into polyads and described by a common effective Hamiltonian that incorporates the relevant coupling interactions. In the polyad-based treatment, fundamental bands and associated hot bands can therefore be handled within the same theoretical framework, allowing resonance interactions between the contributing states to be naturally accounted for. Such a global description is particularly advantageous as it allows to: (i) simulate broadband spectra of polyatomic molecules, where overlapping bands and vibrational couplings can significantly complicate the measured spectra, (ii) compute absolute line strengths of rovibrational transitions, whereas the empirical PGOPHER approach typically provides relative intensities.

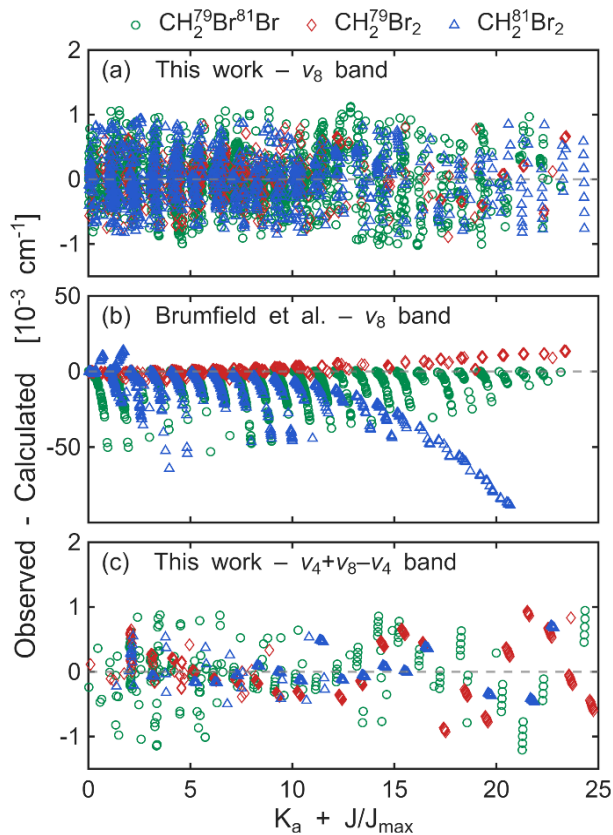

**Fig. 6** Residuals between the observed transition wavenumbers and those from a PGOPHER fit (Obs - Calc) as a function of the upper state $K$ and $J$ quantum numbers of the $\nu_8$ band for (a) our work and (b) using the band parameters of Brumfield et al,[26] and (c) of the $\nu_4+\nu_8-\nu_4$ band for the three isotopologues: CH$_2^{79}$Br$^{81}$Br (green), CH$_2^{79}$Br$_2$ (red), and CH$_2^{81}$Br$_2$ (blue) from our work.

Fig. 7 shows the calculated rovibrational line strengths of the fundamental $\nu_8$ transition polyad (P$_0 \rightarrow$ P$_7$) and the corresponding hot band $\nu_4+\nu_8-\nu_4$ polyad transitions (P$_1 \rightarrow$ P$_8$) for the three isotopologues. The remaining seven and ten vibrational bands in the *ab initio* P$_7$ and P$_8$ models, respectively, were not observed in the experimental spectra, and thus were not assigned, although they may interact with $\nu_8$ and $\nu_4 + \nu_8$. The intensities shown in this figure already account for the natural isotopic abundances of $^{79}$Br and $^{81}$Br (0.5069 and 0.4931, respectively).[35] Here, the nuclear spin statistical weight of the C$_{2v}$ and C$_s$ species are 9:7 and 16:16 for the $A:B$ and $A':A''$ symmetries, respectively. They were used to compute the values of the partition functions given by 2243730 and 4557407 for the C$_{2v}$ and C$_s$ species, respectively. For the mixed isotopologue, the peak intensity is 4.54 × 10$^{-21}$ cm$^{-1}$/(molecule·cm$^{-2}$), which is almost twice as large as the light and heavy isotopologues, with maxima of 2.65 × 10$^{-21}$ and 2.47 × 10$^{-21}$ cm$^{-1}$/(molecule·cm$^{-2}$), respectively. The corresponding P$_1 \rightarrow$ P$_8$ hot band intensities are 1.97 × 10$^{-21}$, 1.15 × 10$^{-21}$, and 1.1 × 10$^{-21}$ cm$^{-1}$/(molecule·cm$^{-2}$) for the mixed, light, and heavy isotopologues, respectively, representing ~43% (on average) of the fundamental band intensities. This is in a good agreement with the simple Boltzmann prediction of ~44% of population distributions at room temperature. This also indicates the significance of polyad interactions in the *ab initio* effective Hamiltonian approach, which captures couplings and predicts the intensities of the different vibrational bands accurately.

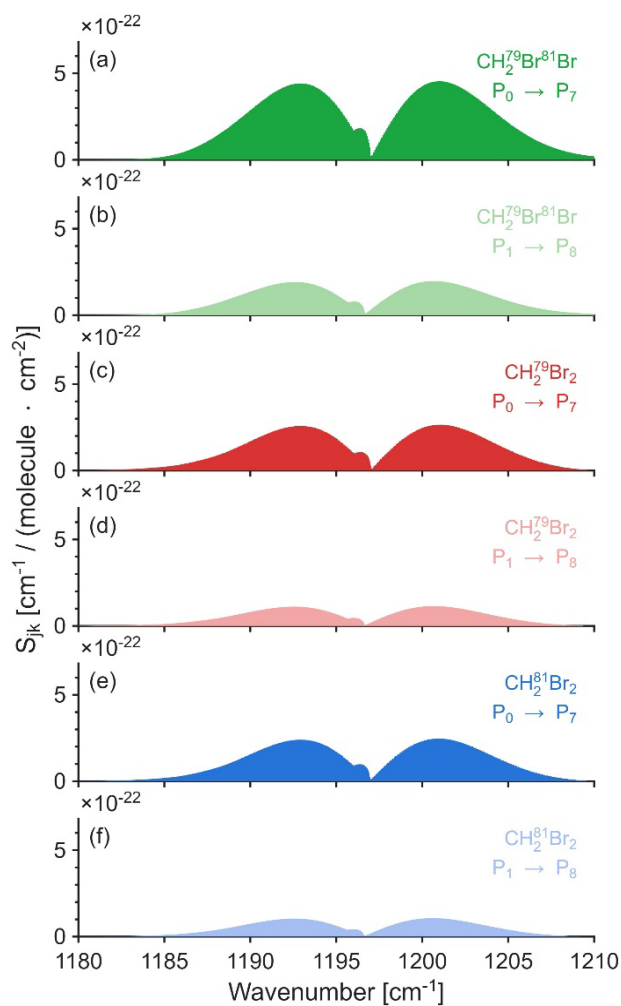

**Fig. 7** Calculated *ab initio*-based intensities of the fundamental ($P_0 \rightarrow P_7$) polyad rovibrational transitions and the hot band ($P_1 \rightarrow P_8$) transitions in the 1180 to 1210 cm$^{-1}$ spectral region for the three isotopologues of dibromomethane: $CH_2^{79}Br^{81}Br$ [green – panels (a) and (b)], $CH_2^{79}Br_2$ [red – panels (c) and (d)], and $CH_2^{79}Br_2$ [blue – panels (e) and (f)].

Fig. 8 presents the differences in line positions between empirically assigned transitions (Obs) and the *ab initio*-based effective Hamiltonian calculations (Calc) for the $\nu_8$ and $\nu_4+\nu_8-\nu_4$ bands of the mixed isotopologue. The majority of transitions are reproduced with small residuals, resulting in RMS values of 0.008 cm$^{-1}$ for the 2007 $\nu_8$ transitions up to $J_{max}$ = 80 in in panel (a), and 0.007 cm$^{-1}$ for the 365 transitions of $\nu_4+\nu_8-\nu_4$ shown in panel (b). These values are, however, about 22 times larger than those obtained from empirical fit (see Table 1).

A distinct 'anomaly' is observed around $K_a$ = 16 and 17, where residuals reach 0.07 and 0.05 cm$^{-1}$, respectively. This anomaly was not observable in the empirical fit (see Fig. 6(a)), as it is likely absorbed by the floating of rotational and centrifugal distortion constants in PGOPHER analysis. However, for high-$J$ transitions of $K_a$ = 16 and 17 that were not included in the empirical, this anomaly can become visible, as it can no longer be compensated by the fitted parameters. Excluding transitions with |Obs–Calc| > 0.02 cm$^{-1}$ in Fig. 8(a) results in an RMS value of 0.003 cm$^{-1}$.

Such anomaly in Fig. 8(a) could be possibly attributed to the fact that the *ab initio* model incorporates numerous diagonal and non-diagonal parameters associated with vibrational bands that are not experimentally observed. These parameters were therefore fixed to their *ab initio* values, which are likely not sufficiently accurate. For example, the trend observed in Fig. 8 appears correlated with the Coriolis coupling between the $\nu_8$ and $\nu_3+\nu_9$ bands. The maximum contribution of this coupling to the $\nu_8$ = 1 energy levels is ~0.5 cm$^{-1}$, at $K_a$=16. One possible route to improve the quality of the *ab initio* effective coupling parameters is first to enhance the accuracy of the PES by optimizing certain force constants. In Fig. 8(b), several outliers in the $\nu_4+\nu_8-\nu_4$ hot band do not follow the overall trend and may reflect misassignments of these weak transitions.

This behaviour reflects the distinct treatment of molecular interactions in the two approaches. In the empirical PGOPHER analysis, the experimental spectrum is reproduced by adjusting rotational constants and centrifugal distortion parameters, resulting in effective spectroscopic constants that implicitly account for weak perturbations between interacting states. In contrast, the *ab initio* - based effective Hamiltonian explicitly treats such interactions within the polyad framework, providing a more detailed picture of the molecular interactions. Therefore these large residuals around $K_a$ = 6 and 17 may also indicate a local coupling that is not yet accounted for in the *ab initio*-based effective Hamiltonian.

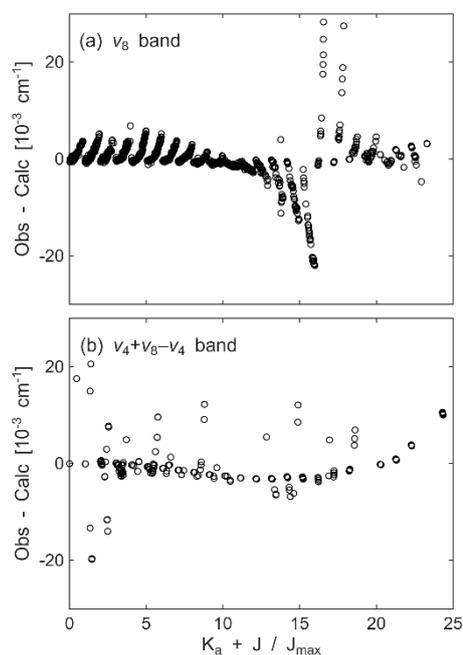

**Fig. 8** (a) Residuals of the transition wavenumber (Obs − Calc) of $CH_2^{79}Br^{81}Br$ from *ab initio*-based effective Hamiltonian calculations as a function of the upper state $K_a$ and $J$ quantum numbers: (a) the $\nu_8$ band, comprising 2007 transitions up to $J_{max}$ = 80; (b) the $\nu_4+\nu_8-\nu_4$ band, comprising 365 transitions up to $J_{max}$ = 80.

The calculated *ab initio*-based line intensities were subsequently used to determine the absorption cross-sections and compared with the experimental data. Fig. 9(a) shows the simulated absorption cross-sections together with PNNL reference spectrum (black). The *ab initio* spectra were simulated using a Voigt line-shape function, with the self-pressure broadening being almost negligible at the measurement pressure of 31 μbar, and set to a value of 0.2 cm$^{-1}$·atm$^{-1}$, typical for halogenated molecules.[35] The spectra were then convolved with a *sinc* function to reproduce the instrumental response of the high-resolution comb measurements.

The contributions of the different polyads are also shown. In this spectral region, the dominant contribution arises from the $P_0{\rightarrow}P_7$ polyad (cyan), mainly associated with the fundamental $\nu_8$ band, accounting for ~50% of the total cross-section. Inclusion of higher polyads, dominated by hot-band transitions originating from the low-lying $\nu_4$ mode, gradually improves the agreement with experiment. Specifically, adding the $P_1{\rightarrow}P_8$ polyad (green), dominated by the $\nu_4+\nu_8-\nu_4$ hot band, increases the agreement to 72%. Further inclusion of the $P_2{\rightarrow}P_9$ (blue) and $P_3{\rightarrow}P_{10}$ (red) polyads, dominated by $2\nu_4+\nu_8-2\nu_4$ and $3\nu_4+\nu_8-3\nu_4$ transitions, respectively, improves the agreement to 83 %.

The remaining discrepancy (17%) is attributed to contributions from higher polyads, e.g., $P_4{\rightarrow}P_{11}$, involving transitions such as $\nu_9+\nu_8-\nu_9$ and $(\nu_3+\nu_4)+\nu_8-(\nu_3+\nu_4)$, which are not included due to insufficient accuracy of their band centres. Inclusion of these bands without further refinement leads to distortions in the simulated spectral contour.

The lower panel of Fig. 9(a) shows the relative difference in cross-section ($\Delta\sigma$, in %) between the simulated and PNNL spectra. The largest discrepancies are observed in the band wings, indicating missing contributions from higher-order hot bands and limitations in the current model, particularly at high-$J$ values.

Apart from the remaining higher polyads, several factors may contribute to the remaining discrepancies. (i) Uncertainties in the PNNL reference data; according to the metadata, a small rescaling of 1.016 was applied to account for impurities. (ii) The existence of a nearby weak bands in the lower wavenumber range that is not included in the present simulations of Fig. 9. (iii) Limitations of the theoretical model; the dipole moment surface used in here was constructed for the CH$_2^{79}$Br$_2$ isotopologue, whereas the mixed isotopologue dominates experimentally, and the treatment of isotopic effects on transition frequencies and intensities remain a challenge for *ab initio* calculations. Finaly, (iv) the completeness of the line list is limited by the convergence with respect to rotational excitation and hot-band contributions. For a heavy molecule such as CH$_2$Br$_2$, the density of states increases rapidly with energy, making it computationally demanding to include all relevant transitions. The combined variational-effective Hamiltonian approach employed here provides a practical solution, but the polyad formalism becomes less reliable at high-$J$, where interactions between states become increasingly complex. A more complete treatment will be presented in an accompanying theoretical study, aiming to provide *ab initio*-based line positions and intensities up to 2000 cm$^{-1}$.

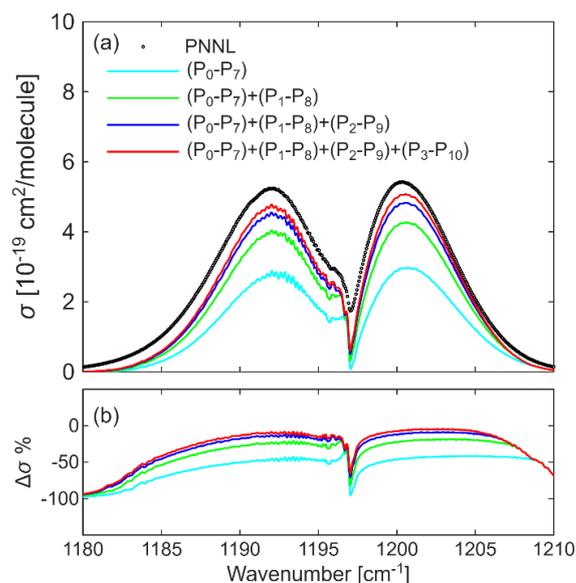

**Fig. 9** (a) Absorption cross-section, $\sigma$, of CH$_2$Br$_2$ at room temperature simulated from *ab initio*-based effective Hamiltonian model, including successive contribution from ($P_0{\rightarrow}P_7$), ($P_0{\rightarrow}P_7$)+($P_1{\rightarrow}P_8$), ($P_0{\rightarrow}P_7$)+($P_1{\rightarrow}P_8$)+($P_2{\rightarrow}P_9$), and ($P_0{\rightarrow}P_7$)+($P_1{\rightarrow}P_8$)+($P_2{\rightarrow}P_9$)+($P_3{\rightarrow}P_{10}$), compared with the PNNL database. (b) Relative differences in cross-section ($\Delta\sigma$, %) between the simulated and PNNL spectra.

## Conclusions

We report the first high-resolution absorption cross-section of CH$_2$Br$_2$ in the long-wave mid-infrared region from 1180 to 1210 cm$^{-1}$. This spectral range is dominated by the intense CH$_2$ wagging vibration ($\nu_8$ band), which is ∼50 stronger than the C−H stretch vibrations in the 3.3 μm region. This makes it particularly suitable for spectroscopic detection of CH$_2$Br$_2$ in applications such as workplace safety monitoring, where it can occur as a by-product in treated ballast water, as well as in environmental sensing. The reported cross-section shows an overall good agreement across the entire $\nu_8$ band region with the low-resolution FT-IR measurements of the PNNL database.

A spectroscopic model of measured spectrum is developed for this asymmetric top molecule based on an empirical fitting, as implemented in PGOPHER. The model explicitly accounts for the natural isotopic abundances of $^{79}$Br and $^{81}$Br, and for the presence of $\nu_4+\nu_8-\nu_4$ hot bands cross-interfering with the fundamental $\nu_8$ bands. Band parameters (origin, rotational constants, and centrifugal distortion constants) of both the $\nu_8$ fundamental and the $\nu_4+\nu_8-\nu_4$ hot bands were determined for all three isotopologues. Compared to previously reported band parameters retrieved from a fit to narrowband (1.78 cm$^{-1}$) supersonically cooled spectra,[25, 26] the present broadband analysis provides global agreement between measured and simulated spectra across a wide range of rotational quantum numbers. A least squares fit of 3034 unique frequency assignments yields a root-mean-square error of 0.00037 cm$^{-1}$ (or 11.1 MHz − average of the six bands). These assigned transitions were used to refine a non-empirical, *ab initio* based effective Hamiltonian for calculating a first complete line list of CH$_2$Br$_2$ isotopologues. Successive inclusion of the dominant

polyads – P$_0\to$P$_7$, P$_1\to$P$_8$, P$_2\to$P$_9$, and P$_3\to$P$_{10}$ – corresponding mainly to the $\nu_8$, the $\nu_4+\nu_8-\nu_4$, $2\nu_4+\nu_8-2\nu_4$ and $3\nu_4+\nu_8-3\nu_4$ bands, respectively, yields agreement with PNNL cross-section at the ~83% level. The remaining discrepancy is attributed primarily to the higher order polyads, e.g., P$_4\to$P$_{11}$ involving bands such as $\nu_9+\nu_8-\nu_9$ and $(\nu_3+\nu_4)+\nu_8-(\nu_3+\nu_4)$, which are not yet included due to insufficient accuracy of their band centres.

Overall, this study provides accurate high-resolution absorption cross-section data in the 8.35 μm spectral region, together with spectroscopic models for all three isotopologues of CH$_2$Br$_2$ in the CH$_2$ wagging vibrational region. The reliability of these models at high-$J$ values still need to be validated against cooled samples through broadband measurements, ideally at variable temperatures. Nevertheless, these results provide the necessary spectroscopic data (when relying on the strong peaks around the band centres) for optical monitoring of CH$_2$Br$_2$ in workplace and some environmental settings, particularly in harbour and ballast-water treatment contexts. In addition, it could the enable quantitative assessments of the detectability of CH$_2$Br$_2$ in planetary atmospheres.

## Author Contributions

**Ibrahim Sadiek:** Conceptualization, Formal analysis, Visualization, Project administration, Writing - Original draft preparation; **Aleksandr Balashov**: Investigation, Formal analysis, Software, Visualization, Writing - Review & Editing; **Adrian Hjältén**: Investigation, Formal analysis, Visualization, Writing - Original draft preparation; **Michael Rey**: Formal analysis, Software, Validation, Writing - Review & Editing; **Oleg Egorov:** Formal analysis, Writing - Review & Editing; **Aleksandra Foltynowicz:** Conceptualization, Supervision, Funding acquisition, Resources, Writing - Review & Editing.

## Conflicts of interest

There are no conflicts to declare.

## Acknowledgments

A. Foltynowicz thanks Grzegorz Soboń for the loan of the optical frequency comb source. This work was supported by funding from the Knut and Alice Wallenberg Foundation (KAW 2020.0303), the Swedish Research Council (2020-00238) and the Kempe foundation (JCSMK24-0034). I. Sadiek would like to acknowledge support from the chair of Experimental Physics V at Ruhr-University Bochum.